\documentclass[conference]{IEEEtran}
\IEEEoverridecommandlockouts
\usepackage{cite}
\usepackage[normalem]{ulem}
\usepackage{amsmath,amssymb,amsfonts}
\usepackage{algorithmic}
\usepackage{subcaption}
\usepackage{multirow}
\usepackage{graphicx}
\usepackage{textcomp}
\newcommand{\degree}{$^\circ$}
\usepackage{xcolor}
\def\BibTeX{{\rm B\kern-.05em{\sc i\kern-.025em b}\kern-.08em
    T\kern-.1667em\lower.7ex\hbox{E}\kern-.125emX}}
\usepackage{fancyhdr}
\fancypagestyle{AcceptanceSentence}{%
  \lhead{This work has been accepted for publication in the IEEE GLOBECOM 2026 Conference.}
}

\begin{document}

\title{The Price of the Golden 6G Band: Evaluation of Beam Management Effort in FR3}

\author{\IEEEauthorblockN{Clémence Altmeyerhenzien, Ljiljana Simić, Marina Petrova}
\IEEEauthorblockA{\textit{Mobile Communications and Computing, RWTH Aachen University}, Germany \\
Email: \{clemence.altmeyerhenzien, simic, petrova\}@mcc.rwth-aachen.de}
}

\maketitle
\thispagestyle{AcceptanceSentence}

\begin{abstract}
Frequency Range 3 (FR3), 7.125–24.25~GHz, regarded as the “golden band” for 6G networks, has less challenging propagation characteristics than FR2 while offering much wider bandwidth for high data rate applications than FR1. Reusing existing FR1 infrastructure for FR3 network deployments requires gNodeBs (gNBs) to employ antenna arrays and perform beam management, which has proven challenging at FR2. In this paper, we extensively study and characterize the beam management effort in an FR3 urban network, in terms of: beam alignment sensitivity, number of directional link opportunities,  gNB handover and beam switch  rates, and beam steering distance. Our results show that achieving a high and stable mobile throughput requires significant beam management effort across FR3 bands. While the beam tracking requirements are less stringent at the lower frequencies due to wider beams, the beam switching rate to a non-adjacent beam is relatively comparable at FR3 and FR2. 
\end{abstract}


\section{Introduction}

Following the opening of the upper-6GHz band (6.425–7.125 GHz) to mobile communication, interest in alternatives to FR1 and FR2 spectrum continues to grow in the context of 6G. In particular, the FR3 range (7.125–24.25 GHz), often called the “golden band”, offers an appealing trade-off: it exhibits more favourable propagation characteristics than FR2, while still providing much larger bandwidth than FR1 \cite{testolina2024sharing, cui20256g, ying2025upper, shakya2025urban}. As future cellular networks are expected to carry increasingly high and heterogeneous traffic volumes, including traffic from AI and sensing tasks, achieving high data rates, robustness, and seamless wireless mobile connectivity remains a key 6G target.  

Reusing the network infrastructure of existing 5G NR FR1 deployments in the FR3 band while achieving comparable coverage and mitigating interference with incumbent users necessitates the use of directional antennas \cite{kang2024cellular}. Importantly, this introduces significant overhead due to beam management procedures such as beam acquisition and tracking, to establish and maintain the directional link with a mobile UE—just like in FR2 deployments where beam management for seamless mobile connectivity remains difficult. The FR3 band covers a wide range of frequencies, and as frequency increases, the channel becomes more sparse, making initial access and tracking more challenging and causing significant data rate loss in case of beam misalignment \cite{ichkov2021millimeter}. Consequently, it is unclear how much beam management effort is required across the FR3 band to obtain the achievable data rates under ideal beam alignment, and whether operating in FR3 is actually less complex than FR2. Understanding the beam management effort is thus crucial for successfully exploiting the FR3 band since active directional beam management is essential to provide seamless 6G mobile coverage. It is also key to enabling spectrum agility in 6G \cite{kang2024terrestrial}, since effectively switching between frequency bands to leverage their advantages requires a clear understanding of the associated system-level trade-offs. 

Existing studies on the FR3 band assume ideal beamforming without yet considering the associated beam management effort. The authors in \cite{testolina2024sharing} evaluate throughput and coverage assuming antennas that can be steered in the exact direction of the UEs, while the authors in \cite{kang2024cellular} select beamforming vectors that maximize the gNB-UE link. In practice, the 3GPP initial-access method still relies on an exhaustive search with codebook-based beamforming \cite{3gpp_ts38213}, introducing overhead due to continuous beam adjustment to track mobile users \cite{8458146}. Since beamforming procedures would be inherited from FR2 \cite{8458146}, it remains unclear to what extent FR3 differs from FR2 in terms of beam management complexity. 
The authors in \cite{shakya2025urban, shakya2025angular} report a wider angular spread and a higher number of multipath components for lower FR3, which could offer more directional link opportunities. However, these studies are limited to channel characterization and do not explicitly address the implications for beamforming overhead.  The authors in \cite{cui20256g} study cross-band beamforming similarity, i.e. whether beam scanning in a lower band can assist beamforming in a higher band, suggesting that this could reduce beam management effort. However, \cite{cui20256g} does not quantify whether these gains are sufficient for practical FR3 deployment. In general, it remains unclear whether FR3 sufficiently reduces the beam management effort compared to FR2 to make practical implementation feasible.

\begin{figure*}[!t]
\centering
    \vspace{-4pt}
\end{figure*}
To address this gap, we present the first comprehensive evaluation of beam management effort in FR3. We consider an FR3 cellular network with a gNodeB (gNB) density comparable to FR1 deployments and use ray-tracing in a realistic urban environment to obtain a spatially consistent channel. We assume codebook-based beamforming with the minimum array size per FR3 band to achieve FR1-equivalent coverage. First, we evaluate the achievable signal-to-interference-and-noise ratio (SINR) and data rate performance and study how strict the corresponding beam alignment requirements are at FR3 compared to an FR2 baseline. Second, we quantify the number of valid beams above an SINR threshold, showing that the beam alignment required to achieve good coverage remains rather strict when moving from FR2 to lower FR3 bands. Third, we characterize beam management effort at FR3 for a mobile UE, in terms of the rate of beam-switch and handover events. Our results confirm that the spectrum-rich FR3 bands can easily outperform FR1 mobile data rates. However, achieving high and stable mobile throughput requires significant beam management effort across all FR3 frequency bands. Importantly, while the beam tracking requirements are more relaxed at the lower frequencies due to wider beams, the rate of beam switching to a non-adjacent beam is relatively comparable at FR3 and FR2.

\begin{figure}[!t]
\centering
\begin{subfigure}[t]{0.218\textwidth}
    \centering
    \includegraphics[width=\linewidth]{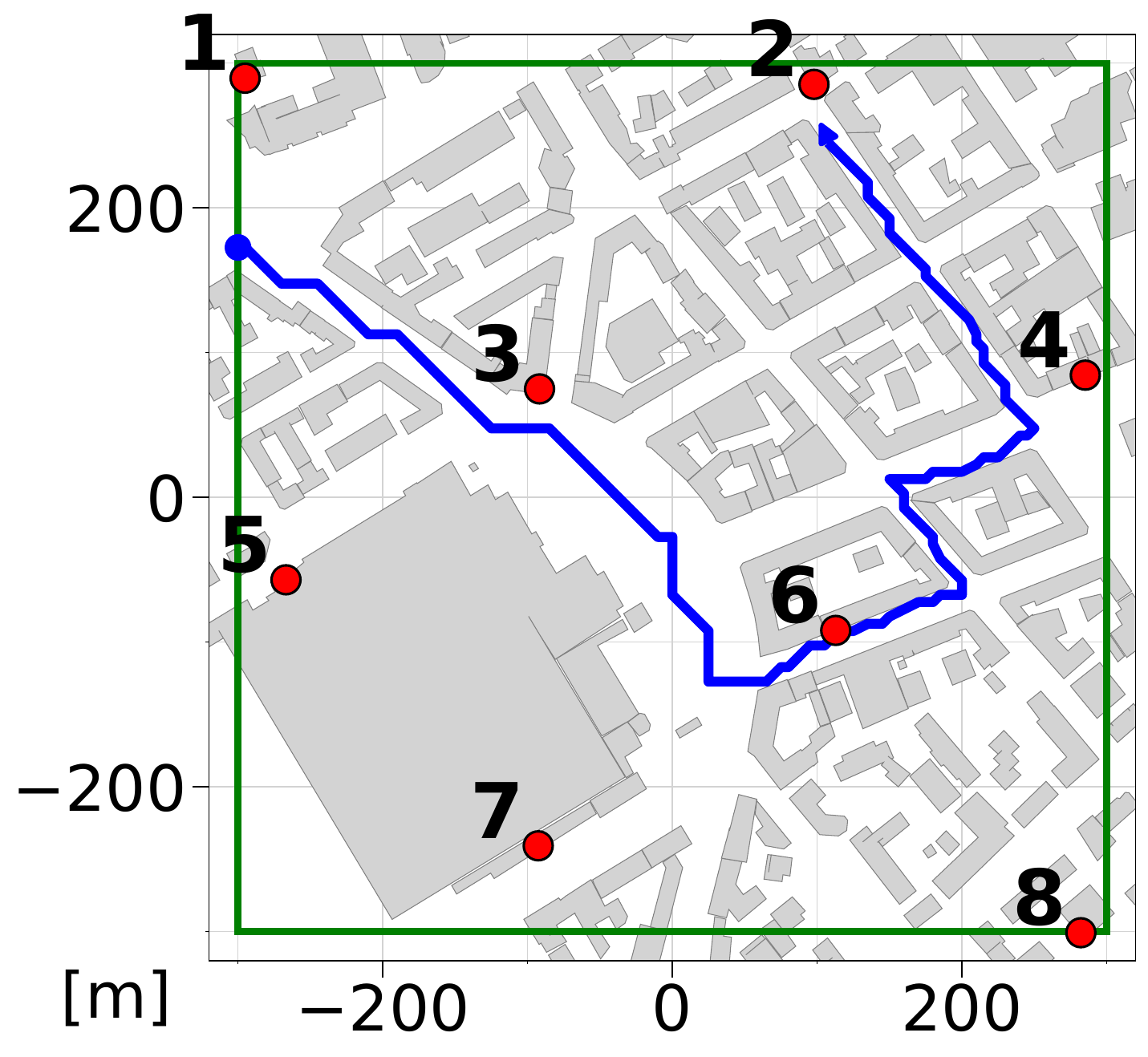}
    \caption{}
    \label{fig:example_path}
\end{subfigure}
\hspace{0.02\textwidth}
\begin{subfigure}[t]{0.186\textwidth}
    \centering
    \includegraphics[width=\linewidth]{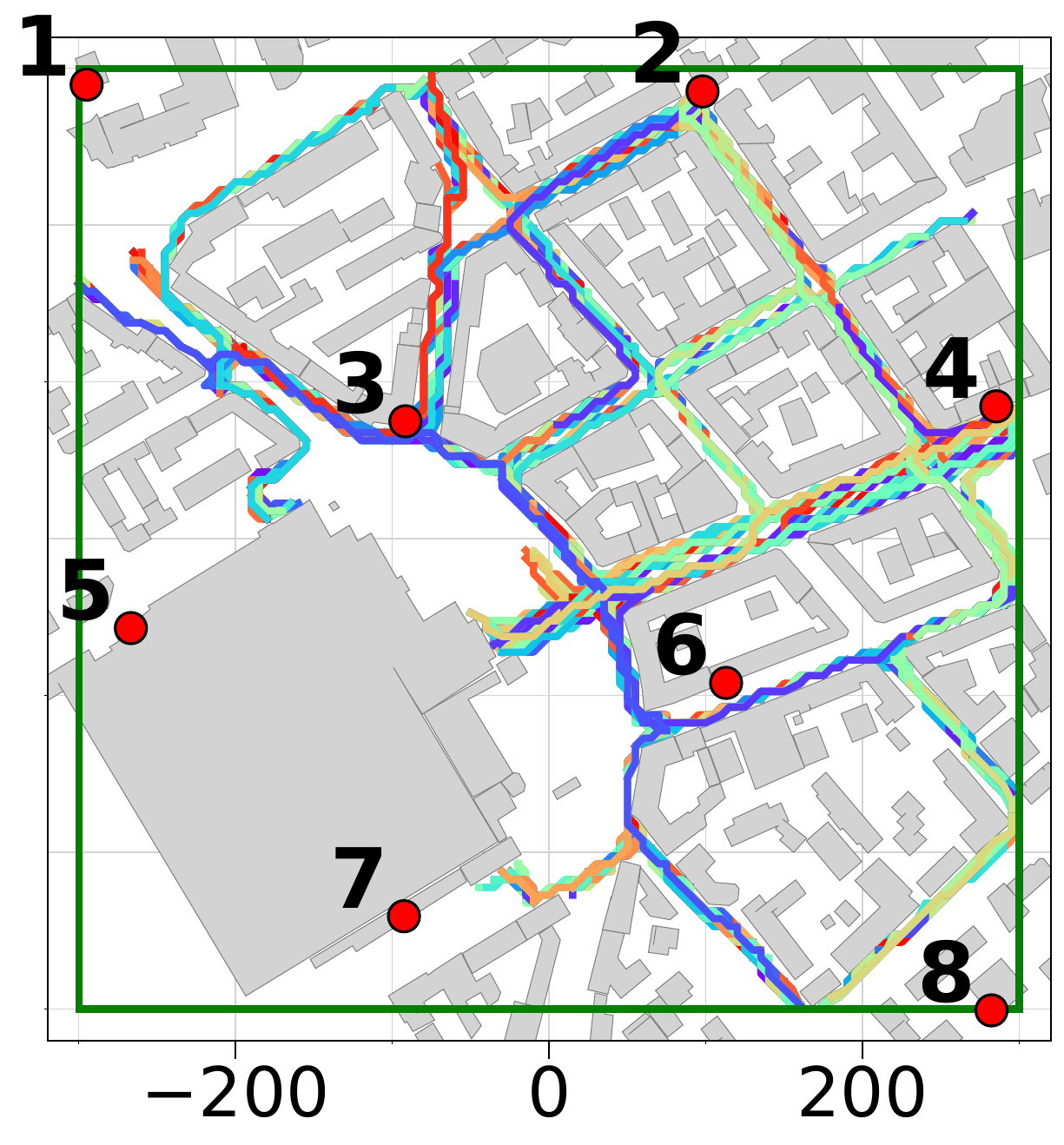}
    \caption{}
    \label{fig:viswalk_paths}
\end{subfigure}
\caption{Network study area (green box) in Frankfurt with gNBs (red dots) with (a) example UE mobility path  (blue) and (b) $M=2000$ pedestrian paths from VisWalk (multi-colour). }
\label{fig:blank_scenario}
\vspace{-3mm}
\end{figure}

\section{System Model}
\label{sec:system_model}

\subsection{Network \& Ray-Tracing Channel Model}
\label{sec:network_raytracing}
We consider a downlink cellular network with $K=8$ gNBs deployed over an area of $800$~m x $800$~m in the city of Frankfurt, in which we select an area of $600$~m x $600$~m to avoid edge effects, as shown in Fig.~\ref{fig:blank_scenario}. The gNBs are distributed corresponding to an FR1 macrocell density of approximately $13$~gNB/km$^2$ and located above rooftops on building corners, at a height of $12$~m above the ground. We study three candidate FR3 carrier frequencies $f_c = \{8, 15, 18\}$~GHz and use the 2.1~GHz at FR1 and 28~GHz at FR2 bands as baselines, with corresponding bandwidths specified in Table~\ref{tab:conductivity}. 
All gNBs have a transmit power of $P_{tx} = 33$~dBm.

We perform ray-tracing simulations using Wireless Insite 3.4.4.13 from Remcom assuming an omnidirectional antenna, which we then post-process in Python to add the directional beamforming gain (\textit{cf.} Section 
\ref{sec:beamforming_postprocessing}). We obtain the urban area model of Frankfurt from Open Street Map and assume concrete as the material for all buildings. Within the study area, we consider outdoor user equipment (UE) positions on a $5$m $\times$ $5$m grid, $2$~m above the ground, resulting in total of $L=6651$ UE locations. We set a ray-tracing receive power threshold of $-250$~dBm and render up to $20$ rays per path with a ray spacing of $0.25\text{\degree}$, subject to $6$ reflections and $1$ diffraction. The omnidirectional ray-tracing outputs the set of possible propagation paths between gNB $k$ and UE location $l$, denoted $\mathcal{J}_{k, l}$. Each path $j\in \mathcal{J}_{k, l}$ includes the angle of arrival (AoA) in the azimuth and elevation, respectively $\{\phi^{A}_j, \theta^{A}_j\}$, the angle of departure (AoD) $\{\phi^{D}_j, \theta^{D}_j\}$, and the path loss $L_{j}$. 
\begin{table}[t]
    \caption{Evaluated frequency bands.}
    \label{tab:conductivity}
    \centering
    \begin{tabular}{|c|c|c|c|c|}
        \hline
        \textbf{Range} & \textbf{3GPP Band} & $\boldsymbol{f_c}$ & \textbf{Channel} $\boldsymbol{B}$& $\boldsymbol{R_{max}}$  \\
        \hline
        FR1     &  n1 & 2.1\,GHz   & 20\,MHz  & 84\,Mbps \\
        FR3     &  n104  & 8\,GHz     & 100\,MHz \cite{3gpp_tr38922}& 422\,Mbps  \\
        FR3      & not specified  & 15\,GHz   & 200\,MHz \cite{3gpp_tr38922} & 844\,Mbps \\
        FR3     & not specified  & 18\,GHz   & 300\,MHz \cite{kang2024cellular} & 1.265\,Gbps \\
        FR2       &  n261 & 28\,GHz   & 400\,MHz & 1.687\,Gbps  \\
        \hline
    \end{tabular}
    \vspace{-0.5cm}
\end{table}
For our mobility case study in Section~\ref{sec:mobility_case_study}, we consider the example path in Fig.~\ref{fig:example_path} with UE locations sampled along our $5$m $\times$ $5$m grid, with location index $0$ and index $173$ shown in Fig.~\ref{fig:example_path} as a blue dot and arrow respectively. We also generate mobility performance statistics by running Monte Carlo simulations over $M=2000$ different UE paths across the grid, as illustrated in Fig.~\ref{fig:viswalk_paths}. Each path consists of $17$ to $117$ UE locations and is generated using the commercial mobility simulator VisWalk, which models realistic pedestrian mobility patterns based on the urban layout in our study area.

\subsection{Beamforming \& Antenna Array Model}
\label{sec:beamforming_postprocessing}
We assume the gNB is equipped with $N_{sector}=3$ sectorized panels, each covering $120$\degree, with a fixed orientation in azimuth $\Phi_p \in \{-120\text{\degree}, 0\text{\degree}, 120\text{\degree}\}$ and the same elevation down-tilt\footnote{We note that some previous studies, e.g. \cite{kang2024cellular} and \cite{testolina2024sharing}, used a downtilt of $-12\text{\degree}$ but considered an area in New York City and Boston with multiple buildings above $200$~m and $100$~m, respectively. In contrast, Frankfurt has only one building above $100$~m. We verified that our result trends remain qualitatively consistent and omit other $\Theta$ values for brevity.
} $\Theta$ for all antennas, which we set to $\Theta = 90\text{\degree}$. Each sector is equipped with a uniform rectangular array (URA) with $N\times N$ antenna elements spaced half a wavelength apart. We model each element following the 3GPP antenna pattern with a maximum gain of $G_{E, max} = 8$~dBi \cite{3gpp38901_2024}. The UE antenna is assumed to be omnidirectional. 

We consider codebook-based beamforming, in line with 3GPP specifications \cite{3gpp_ts38213}, assuming a per-sector codebook $\mathcal{C}_p$. Each codebook contains $n_\phi$ evenly spaced beam directions defined by $(\phi_{c_p}, \theta_{c_p})$ relative to the panel boresight, with $n_\phi=\{3, 5, 10, 21\}$ for URA sizes $\{2\times2$, $4\times4$, $6\times6$, $8\times8\}$, each having a half-power beam width of $\{50.8\text{\degree}, 25.4\text{\degree}, 16.9\text{\degree}, 12.7\text{\degree}\} $ with a maximum gain of $G_{max} = \{14.0, 20.0, 23.6, 26.1\}$~dB, respectively. Note that for the FR1 band, we consider an omnidirectional antenna. The full codebook $\mathcal{C}$ is obtained by replicating $\mathcal{C}_p$ across the $N_{sector}$ panels and has size $|\mathcal{C}| = N_{sector} \cdot n_\phi$. Each codeword $c \in \mathcal{C}$ generates a gain pattern denoted by $G_c(\phi^{D}_j, \theta^{D}_j)$, applied to each propagation path $j$. The received power from gNB $k$ at UE location $l$ thus depends on the codeword $c\in \mathcal{C}$, and is defined as
\begin{equation}
    P_{rx}^{k, l}(c) = P_{tx} \sum_{j \in \mathcal{J}_{k, l}}L_j G_c(\phi^{D}_j, \theta^{D}_j)\text{,} 
\end{equation}
where $L_j$ is the path loss of the $j$-th propagation path obtained via ray-tracing (\textit{cf.} Section \ref{sec:network_raytracing}).

\subsection{Metrics \& Definitions}
\label{sec:metrics}

\subsubsection{SNR}

The signal-to-noise ratio (SNR) given beam $c$ is computed as $SNR_{k, l}(c) = P_{rx}^{k, l}(c)/P_N$, where the noise power is given by the thermal noise over the bandwidth $B$, plus a noise figure of $7$~dB.

\subsubsection{SINR}
Assuming frequency reuse $1$ as typical for 5G NR, interference at UE location $l$ may originate from any of the non-serving gNBs. Since the actual interference depends on which codeword (i.e. beam) the interfering gNB is currently using, we model the average interference by assuming that each interfering gNB selects a beam at random from its codebook, with uniform probability of pointing toward location $l$ \cite{simic2017coverage}, giving the SINR as

\begin{equation}
    SINR_{k, l}(c) = \frac{P^{k, l}_{rx}(c)}{P_N + \frac{1}{|\mathcal{C}|}\sum_{k'\neq k} \sum_{c' \in \mathcal{C}} P^{k', l}_{rx}(c')} \text{. }
\end{equation}

\subsubsection{Data Rate}
We estimate the achievable downlink throughput using the attenuated and truncated Shannon bound
\begin{equation}
    R_{k, l}(c)=B \alpha\min\{\psi_{max}, \log_2 (1 + SINR_{k, l}(c))\}
    \label{eq:rate_snr} \text{, }
\end{equation}
where $\psi_{max}=7.4$ is the maximum spectral efficiency, corresponding to the highest modulation and coding scheme from 3GPP table 5.2.2.1-3 \cite{3gpp_ts38214}. 
The corresponding maximum SINR $\rho_{\max}=22.3$~dB thus truncates the auto-rate function to the maximum achievable data rate $R_{max}$, listed in Table~\ref{tab:conductivity} for each frequency band given its bandwidth $B$. We assume the implementation efficiency $\alpha=0.57$ \cite{mogensen2007lte} and a minimum outage SINR of $-5$~dB, under which the rate is $0$~Mbps.

\subsubsection{Number of Valid Beams}
We define a \textit{valid beam} for a gNB-UE location pair $(k, l)$ relative to threshold $\rho$ as a beam with SINR exceeding $\rho$, and the set of valid beams as 
\begin{equation}
\label{eq:valid_beam}
\mathcal{V}_{k, l}^{\rho} = \{c \in \mathcal{C}: SINR_{k, l}(c) \geq \rho\}
\end{equation}
and the number of valid beams is $|\mathcal{V}_{k, l}^{\rho}|$. 

\subsubsection{Optimal Beam and gNB}
We define the optimal beam and serving gNB $c^*_{k, l} \in \mathcal{C}$, $k^*_{l} \in \{1, ..., K\}$ for UE $l$ as
\begin{equation}
    (c^*_{l},k^*_{l} ) = \arg \max_{c \in \mathcal{C}, k \in \{1, \dots, K\}} SINR_{k, l}(c).
    \label{eq:best_beam}
\end{equation}

\subsubsection{Steering Distance}
We define the \textit{steering distance} between two codewords $c_m, c_n \in \mathcal{C}$ as the circular distance
\begin{equation}
    \Delta_c = \min(|c_m - c_n|, |\mathcal{C}| - |c_m-c_n|).
    \label{eq:steering_distance}
\end{equation}

\section{Results}
\label{sec:results_analysis}

\subsection{Achievable Coverage at FR3 for Optimal Beam Alignment}
\label{sec:achievable_coverage}
\begin{figure}[t]
 \centering
    \includegraphics[width=1.0\linewidth]{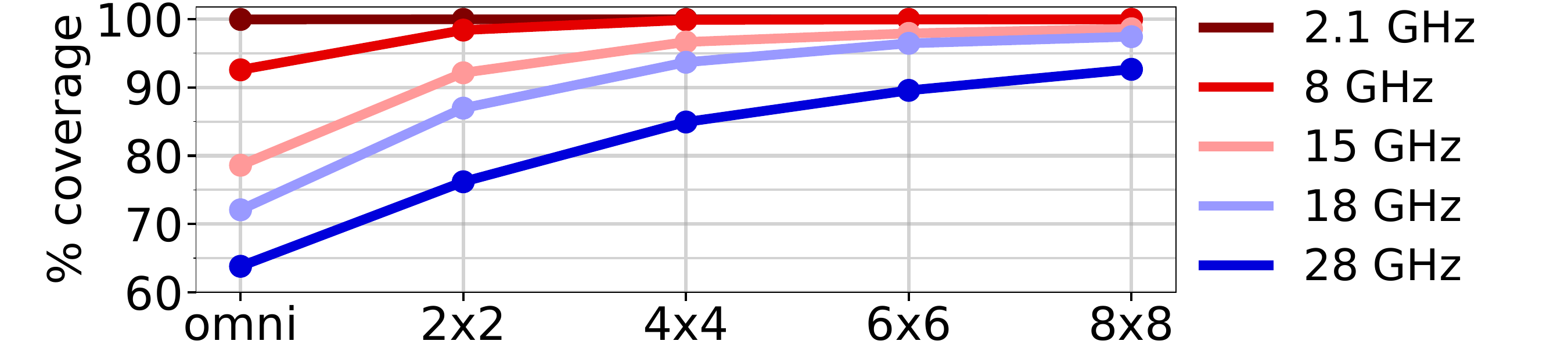}
\caption{Percentage of UE locations covered by $SNR \geq -5$~dB, for different frequency bands and antenna array sizes.}
\label{fig:coverage_summary}
\end{figure}

Let us first establish the minimum required antenna array size at different frequency bands to achieve comparable coverage to FR1, which is our baseline. Fig.~\ref{fig:coverage_summary} presents the percentage of UE locations over the network study area in Fig.~\ref{fig:blank_scenario} covered by an SNR of at least $-5$~dB, for different frequency bands and antenna array sizes, assuming the optimal beam and the best serving gNB at each UE location. Fig.~\ref{fig:coverage_summary} shows that in the FR1 band, all UE locations are covered by an SNR above $-5$~dB. Since the propagation path loss increases with the frequency, achieving comparable coverage to FR1, which we define as $95$\% coverage of network locations, at FR3 requires a $2 \times 2$ array at $8$~GHz, a $4\times4$ array at $15$~GHz, a $6\times6$ array at $18$~GHz (whereas at $28$~GHz a comparable coverage cannot be reached, even with an $8 \times 8$ antenna array). 

\begin{figure}[t]
\centering

\begin{subfigure}[t]{\linewidth}
    \centering
    \includegraphics[width=1\linewidth]{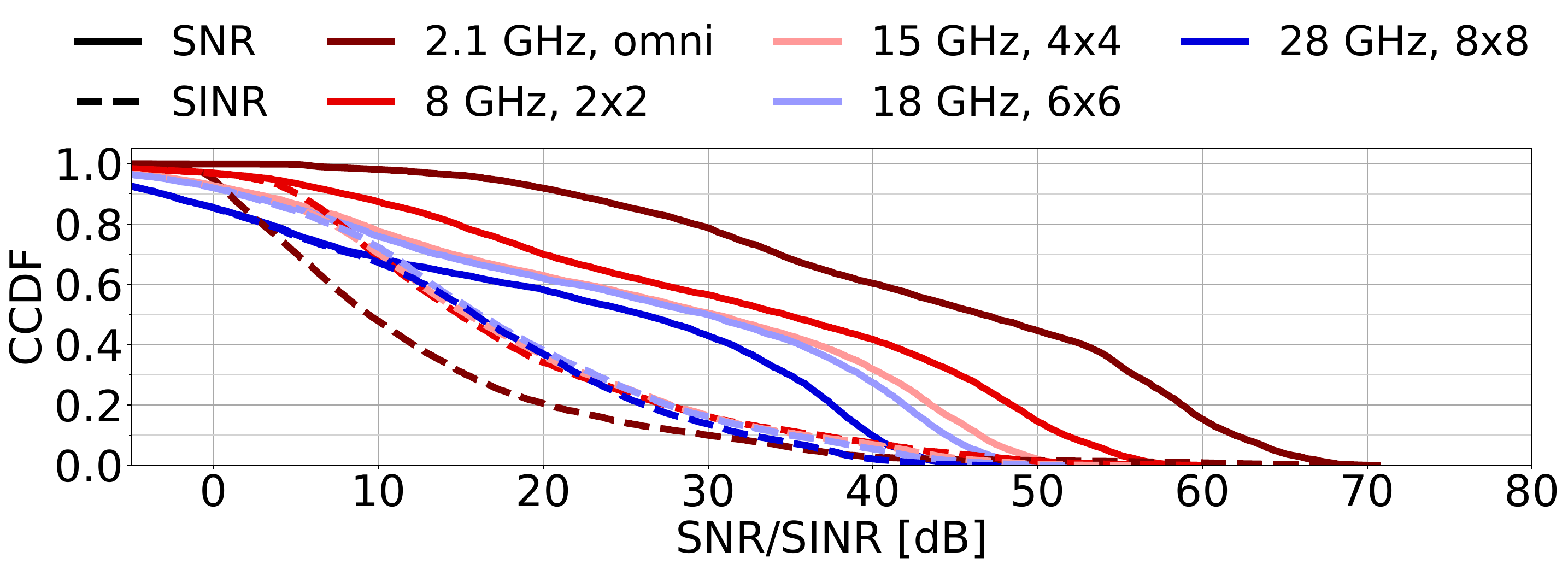}
    \caption{}
    \label{fig:ccdf_snr_90}
\end{subfigure}

\vspace{0.5em}

\begin{subfigure}[t]{\linewidth}
    \centering
    \includegraphics[width=1\linewidth]{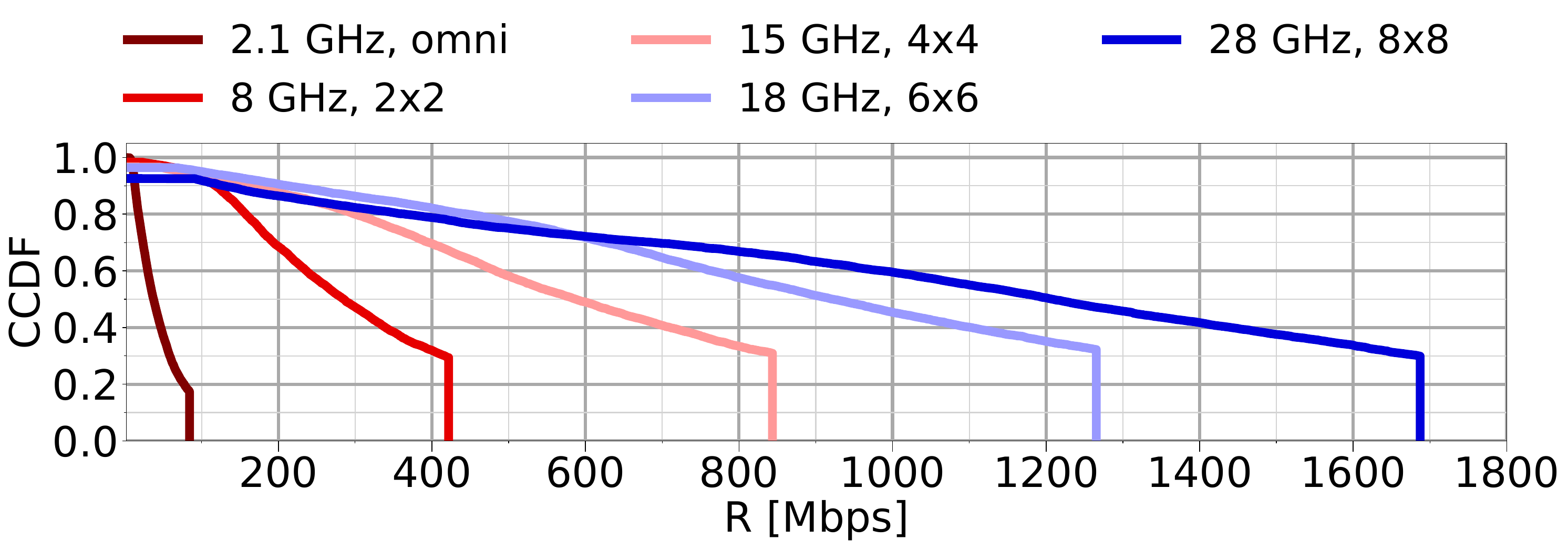}
    \caption{}
    \label{fig:ccdf_dr_90}
\end{subfigure}

\caption{Distribution of the maximum achievable (a) SNR and SINR and (b) data rate over the network for the optimal gNB and beam.}
\label{fig:ccdf_combined}
\vspace{-3mm}
\end{figure}

\begin{figure}
\begin{minipage}{0.95\linewidth}
\begin{subfigure}[t]{0.554\textwidth}
    \centering
    \includegraphics[width=\linewidth]{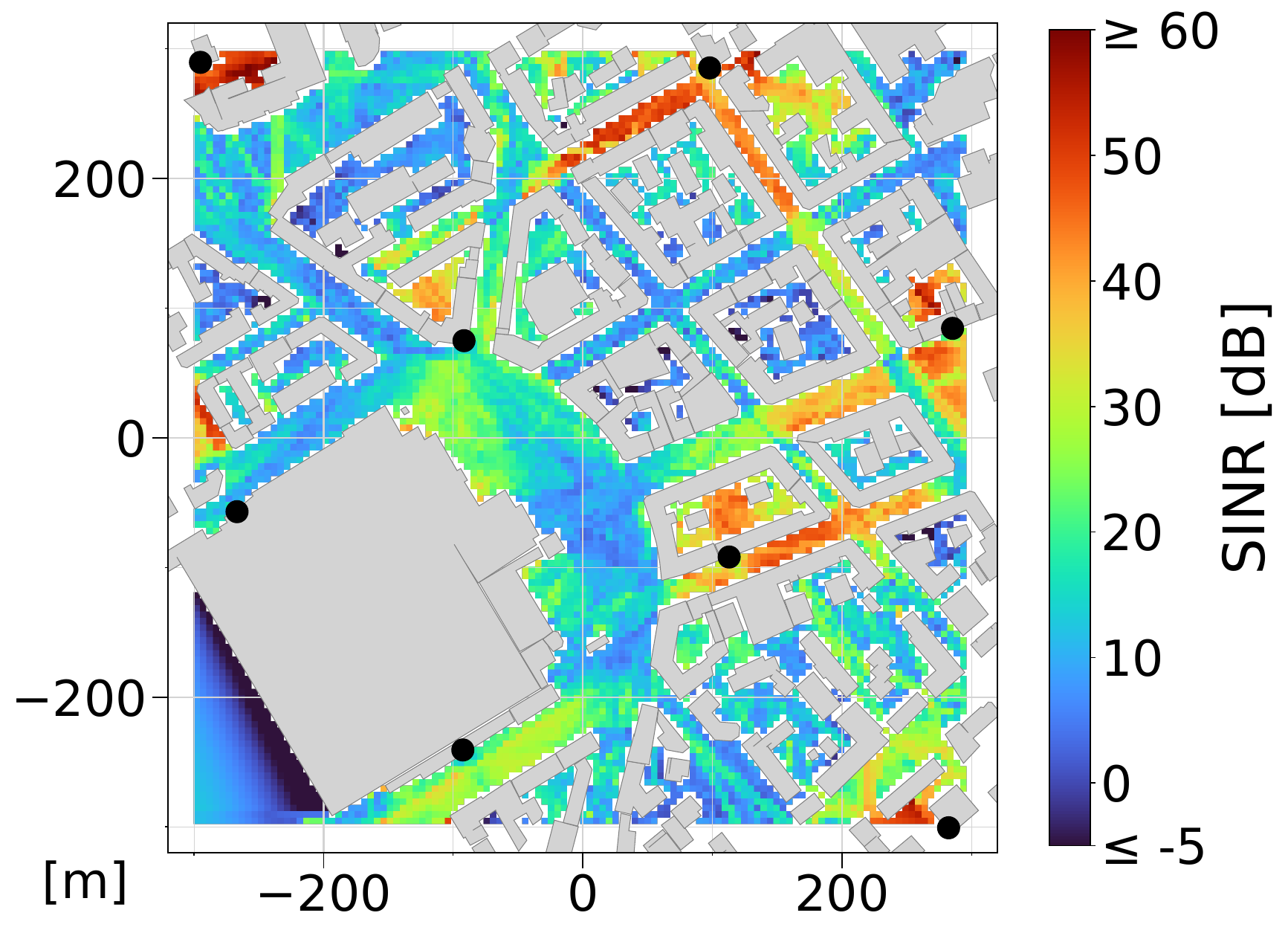}
    \caption{}
    \label{fig:heatmap_sinr_opt_8ghz}
\end{subfigure}
\vspace{0.5em}
\begin{subfigure}[t]{0.374\textwidth}
    \centering
    \includegraphics[width=\linewidth]{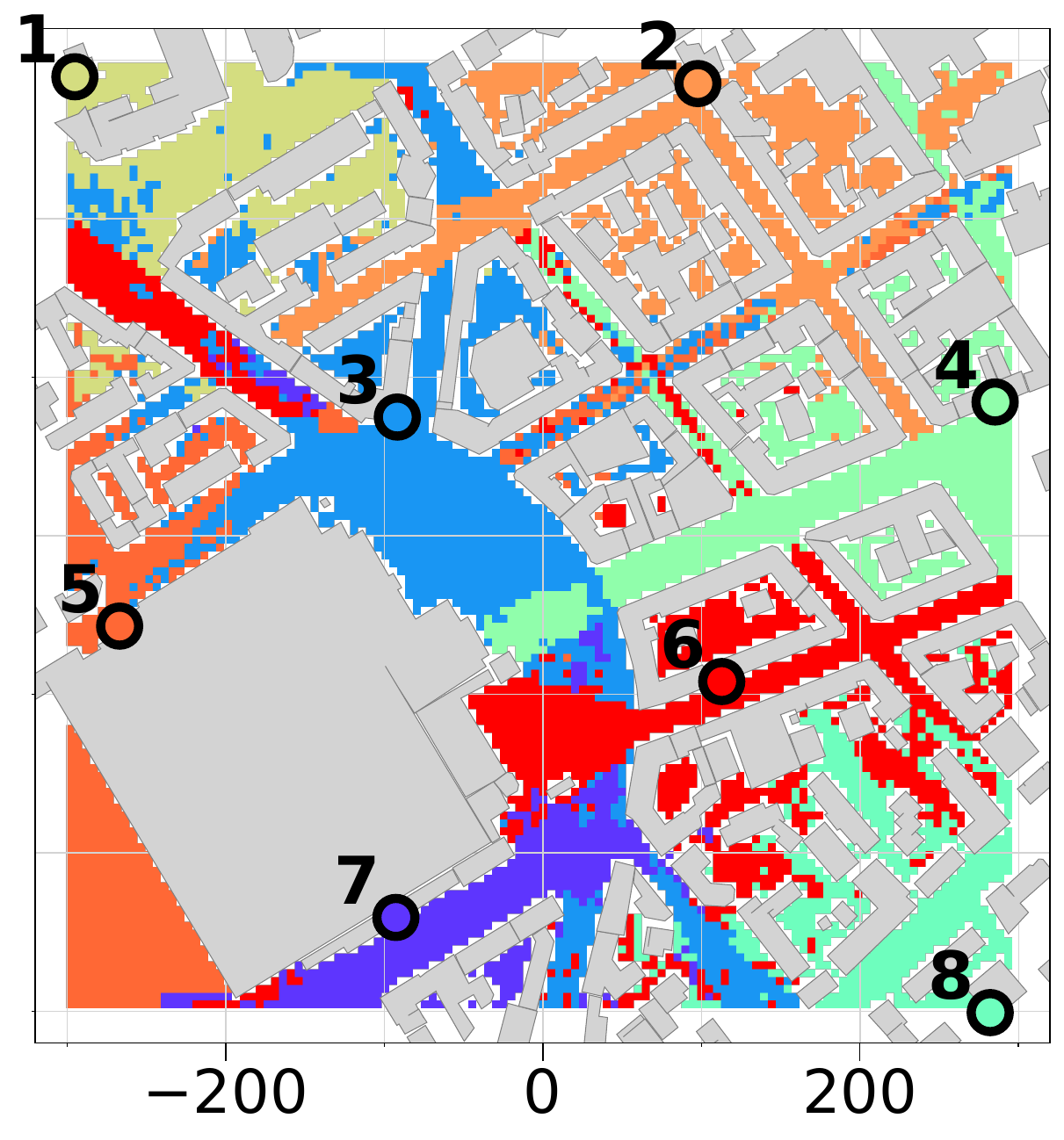}
    \caption{}
    \label{fig:heatmap_best_gNB_8ghz}
\end{subfigure}
\end{minipage}
\caption{Heatmap of (a) maximum achievable SINR in Fig.~\ref{fig:ccdf_snr_90} and (b) best serving gNB $k^*$, for $f_c = 8$~GHz, $2 \times 2$ array.}
\label{fig:heatmap_SINR_servingGnb}
\end{figure}

Fig.~\ref{fig:ccdf_snr_90} presents the distribution of the maximum achievable SNR and SINR for the different bands using these smallest antenna arrays possible to provide FR1-like coverage. Fig.~\ref{fig:ccdf_dr_90} presents the corresponding maximum achievable data rate using (\ref{eq:rate_snr}).  Fig.~\ref{fig:ccdf_snr_90} confirms that decreasing the carrier frequency results in a left-shifted SNR distribution due to lower path loss \cite{kang2024cellular, hu2024channel}. When considering interference, 
the SINR and SNR curves diverge, except for UEs in the noise-dominated region (e.g. bottom 15\textsuperscript{th} percentile of users at 18 GHz). For the remaining users, the SINR distributions at different FR3/FR2 frequencies largely overlap, since the increasing received power 
is undermined by the increasing interference as the frequency decreases. We note that the FR2 and FR3 bands result in a higher SINR (e.g. by $5$~dB for the median user) than the FR1 band due to the benefit of limited spatial interference coming from directional beams versus the FR1 omnidirectional baseline. The spatial distribution over the network of the achievable SINR and serving gNB is similar across frequencies, as illustrated for $8$~GHz in Fig.~\ref{fig:heatmap_SINR_servingGnb}.

Fig.~\ref{fig:ccdf_dr_90} shows that around $18$\% and $30$\% of users in the FR1 and FR3/FR2 networks, respectively, obtain the maximum achievable data rate $R_{max}$ for their respective band (\textit{cf.} Table~\ref{tab:conductivity}). This corresponds to the maximum SINR $\rho_{max}=22.3$~dB of the rate function in (\ref{eq:rate_snr}), achieved by the same proportion of users in Fig.~\ref{fig:ccdf_snr_90}. 
We also observe that the FR3 networks achieve a data rate of over 100 Mbps (which is around the top FR1 throughput) for more than $92$\% of UE locations, thanks to the wide channel bandwidth in these bands. This indicates that FR3 users can obtain the top rate of FR1 in most cases, albeit operating at low spectral efficiency. 

\begin{figure}[t]
\centering

\begin{minipage}{0.47\textwidth} 
\centering

\begin{subfigure}[t]{0.517\textwidth}
    \centering
    \includegraphics[width=\linewidth]{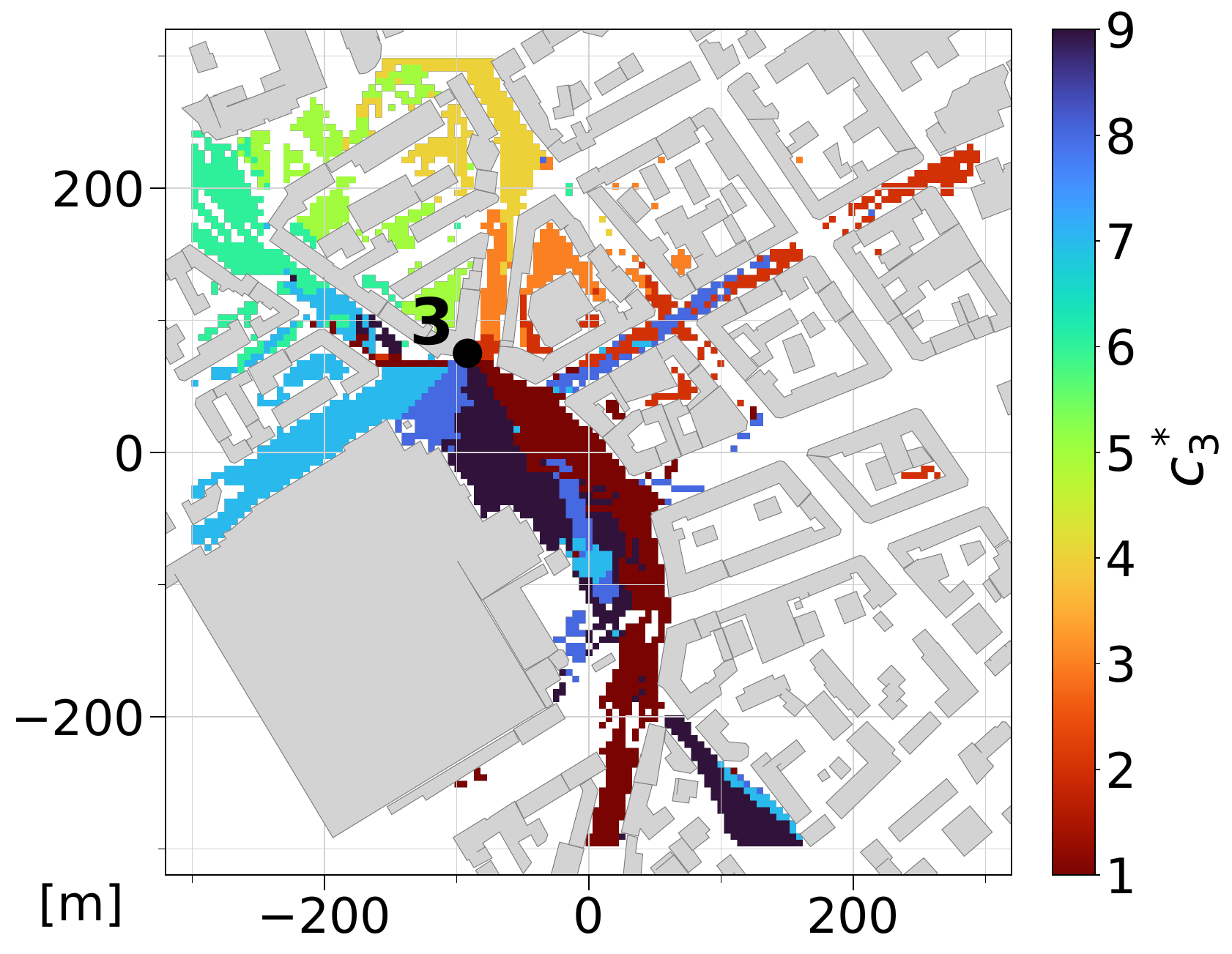}
    \caption{$f_c = 8$ GHz, $2 \times 2$}
    \label{fig:best_beam_2x2}
\end{subfigure}
\hfill
\begin{subfigure}[t]{0.47\textwidth}
    \centering
    \includegraphics[width=\linewidth]{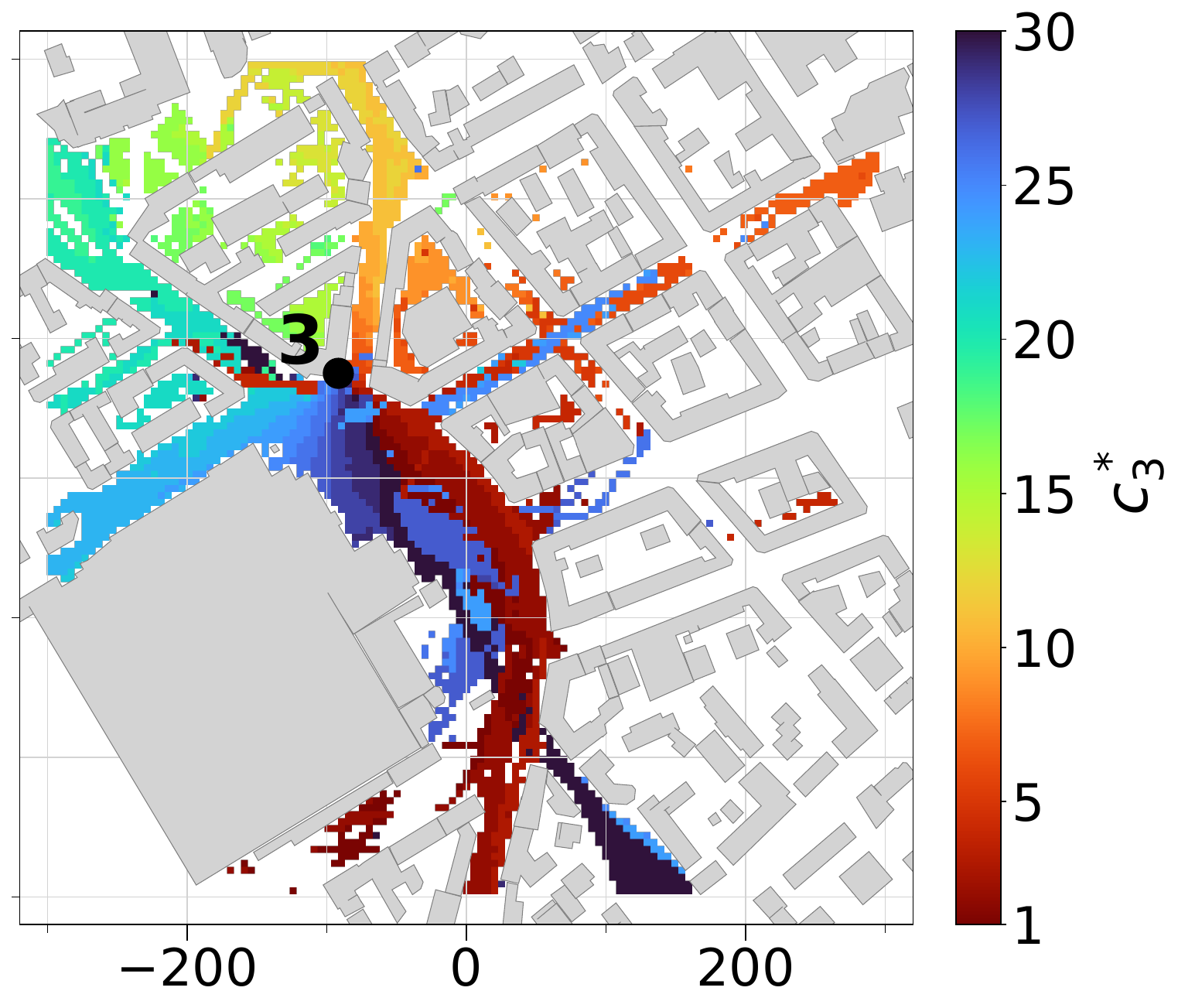}
    \caption{$f_c = 18$ GHz, $6 \times 6$}
    \label{fig:best_beam_6x6}
\end{subfigure}

\end{minipage}

\caption{Heatmap of optimal beam index $c^*$ for gNB $3$ over the network (for UE locations covered by $SINR \geq -5$~dB).}
\label{fig:best_beam}
\end{figure}

Finally, let us examine the nature of the optimal beam alignment underlying the maximum achievable throughput performance in Fig.~\ref{fig:ccdf_dr_90}. Fig.~\ref{fig:best_beam} illustrates the spatial distribution over the network of the optimal beam selection for gNB $3$, i.e. $c^*_{3, l}$  for each UE location $l \in \{1, ..., L\}$, for the 8 GHz ($2 \times 2$ array) and 18 GHz ($6 \times 6$ array) bands. It is evident that the orientation of the optimal beam largely follows the line-of-sight (LoS) direction for UE locations in the open-space area close to the gNB, such that nearby UE locations have adjacent optimal beams, implying straightforward beam tracking. By contrast, in areas where non-LoS coverage dominates, e.g. the narrow street around (100m, 100m), adjacent UE locations have non-adjacent optimal beams, implying more beam management effort. Moreover, Fig.~\ref{fig:best_beam} shows that the size of the network regions where the optimal beam remains the same is smaller for the 18 GHz than for the 8 GHz band, suggesting that maintaining optimal beam alignment for a mobile UE requires more frequent beam switching for higher frequency bands, due to the narrower beams of larger antenna arrays.

\begin{figure}
    \centering
    \includegraphics[width=1.0\linewidth]{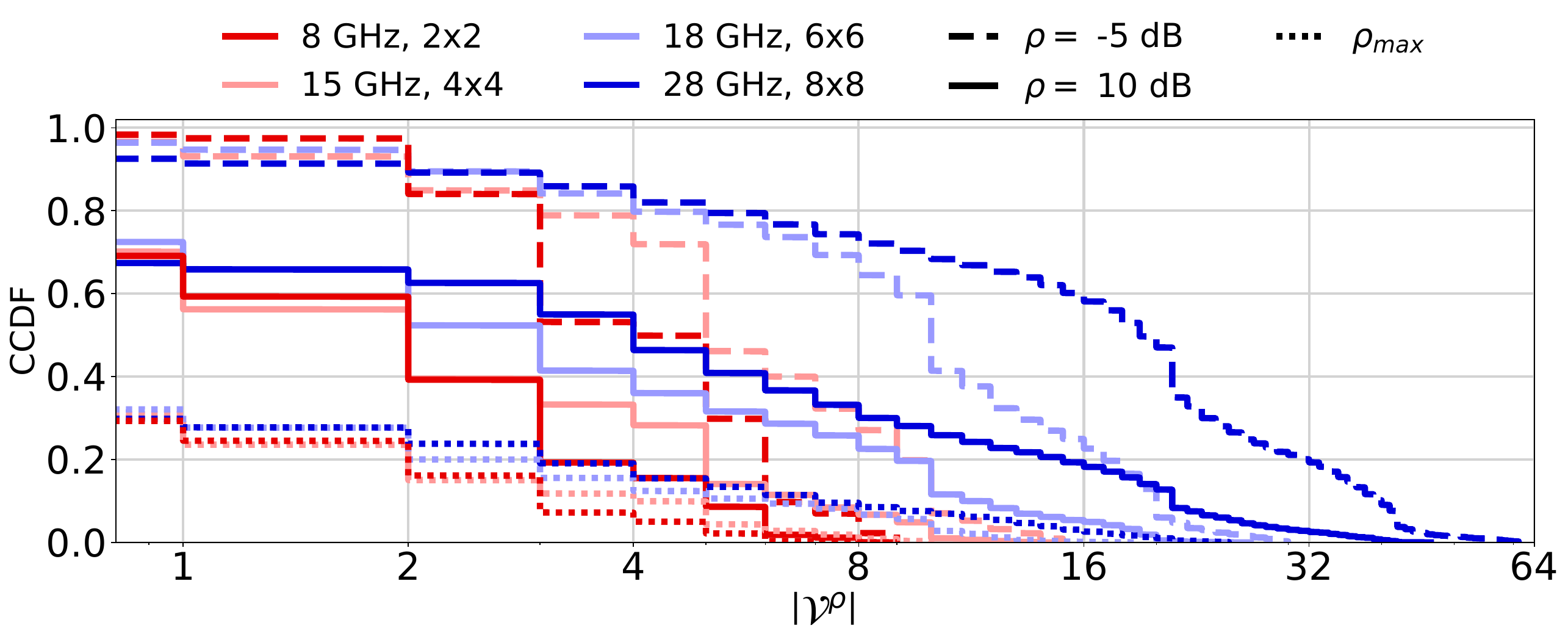}
    \caption{Distribution over the network area of the number of valid beams $|\mathcal{V}^{\rho}|$ at the best serving gNB for different frequency bands and SINR thresholds $\rho$.}
    \label{fig:ccdf_valid_beams_comparison}
    \vspace{-3mm}
\end{figure}

\subsection{Valid Beam Availability at FR3}
\label{sec:valid_beam}
Let us now analyze more generally how strict the beam management requirements are at FR3, by considering the number of directional link opportunities at a given UE location in terms of the number of valid beams for satisfying a target SINR performance, as defined by (\ref{eq:valid_beam}).  Fig.~\ref{fig:ccdf_valid_beams_comparison} presents the distribution over the network of the number of valid beams $|\mathcal{V^{\rho}}|$ at the UE’s best serving gNB $k^*$ to satisfy a minimum SINR of $\rho = \{-5, 10, \rho_{max}\}$, for different FR3 frequencies and FR2 as a baseline. We observe that less than 3\% of UE locations have a fully valid codebook with respect to the outage threshold at the 8 GHz band, reducing to zero as the frequency band increases. Compared to the FR2 baseline, fewer valid beams are available at the lower FR3 frequency bands. However, this difference between FR3 and FR2 reduces as the SINR threshold increases: for $\rho = 10$~dB, the median number of valid beams is 2 at 8~GHz vs. 4 at 28~GHz, whereas the distributions almost overlap for $\rho_{max}$, indicating a comparably strict beam alignment. 
Moreover, although the number of valid beams generally increases in Fig.~\ref{fig:ccdf_valid_beams_comparison} as the frequency increases, the number of valid beams as a \emph{proportion of the whole codebook} decreases. For example, the median number of valid beams satisfying the SINR threshold of $\rho =10$~dB is 2 out of 9 beams (22\%) for the 8~GHz band, 2 out of 15 (13\%) for 15~GHz, 3 out of 30 (10\%) for 18~GHz and 5 out of 63 (8\%) for 28~GHz, suggesting more challenging beam selection at higher frequencies. 

\begin{figure}
\centering
\begin{minipage}{0.46\textwidth}
    \centering
    \begin{subfigure}{\textwidth}
        \centering
        \begin{minipage}{0.522\textwidth}
            \includegraphics[width=\textwidth]{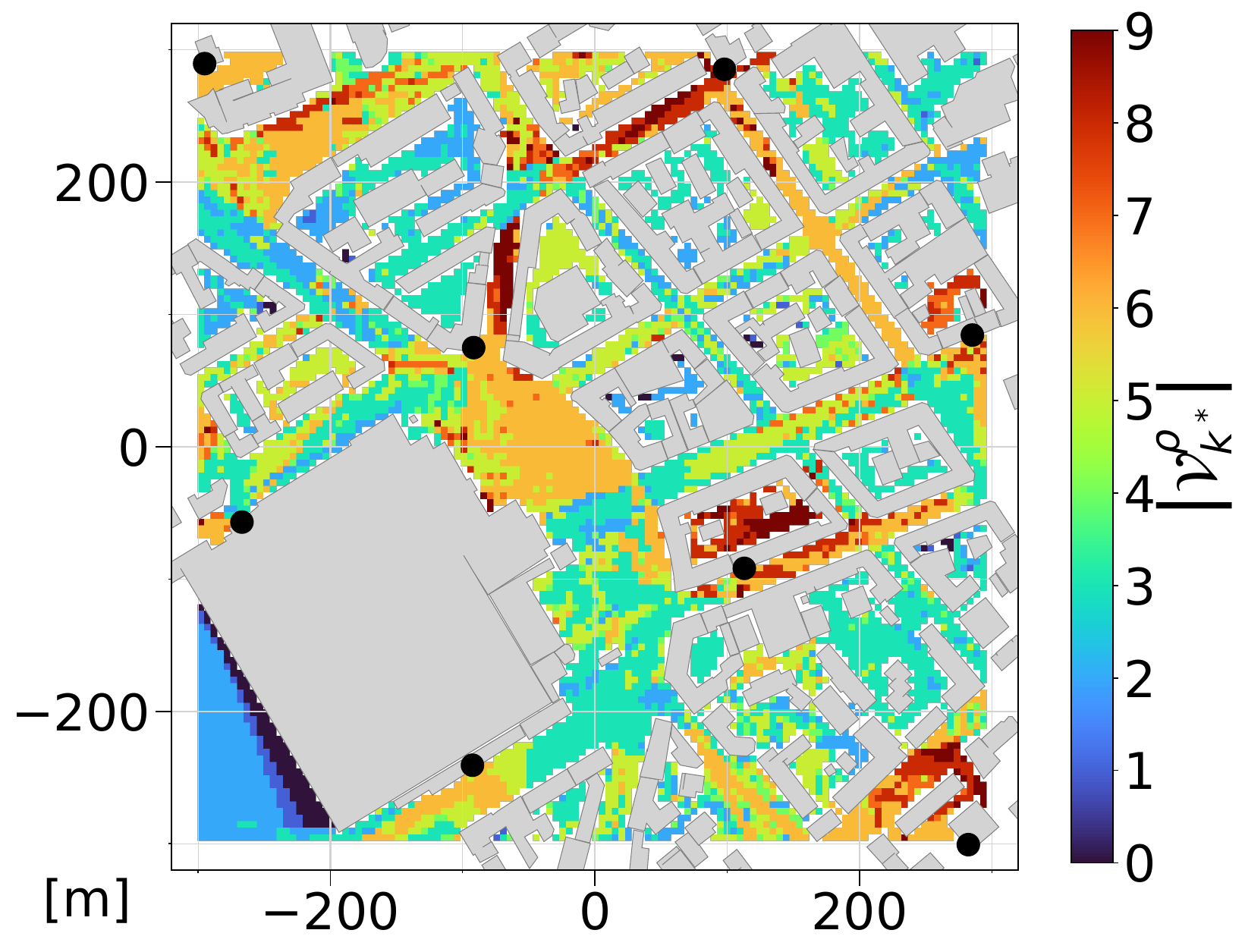}
            \caption{$f_c=8$ GHz, $2 \times 2$}
            \label{fig:valid_beam_8_-5}
        \end{minipage}
        \hfill
        \begin{minipage}{0.462\textwidth}
            \includegraphics[width=\textwidth]{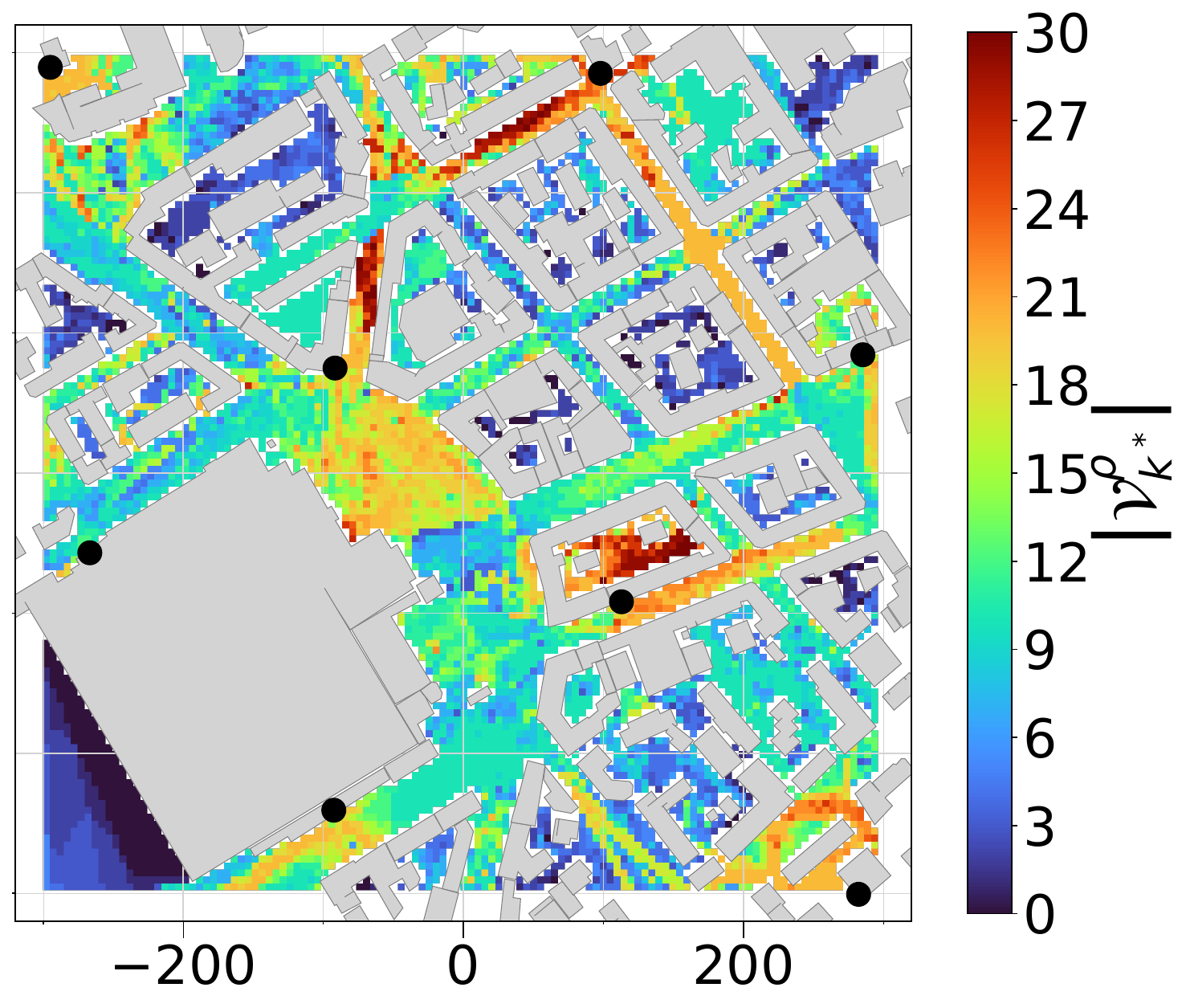}
            \caption{$f_c=18$ GHz, $6\times6$}
            \label{fig:valid_beam_18_-5}
        \end{minipage}
    \end{subfigure}
    \caption{Heatmap of number of valid beams $|\mathcal{V}_{k^*}^{\rho}|$ at the best gNB $k^*$, for different FR3 bands ($\rho = -5$~dB).}
    \label{fig:valid_beam_spatial}
    \vspace{0.4cm}
\end{minipage}
\begin{minipage}{0.43\textwidth}
    \centering
    \begin{subfigure}{\textwidth}
        \centering
        \begin{minipage}{0.46\textwidth}
            \includegraphics[width=\textwidth]{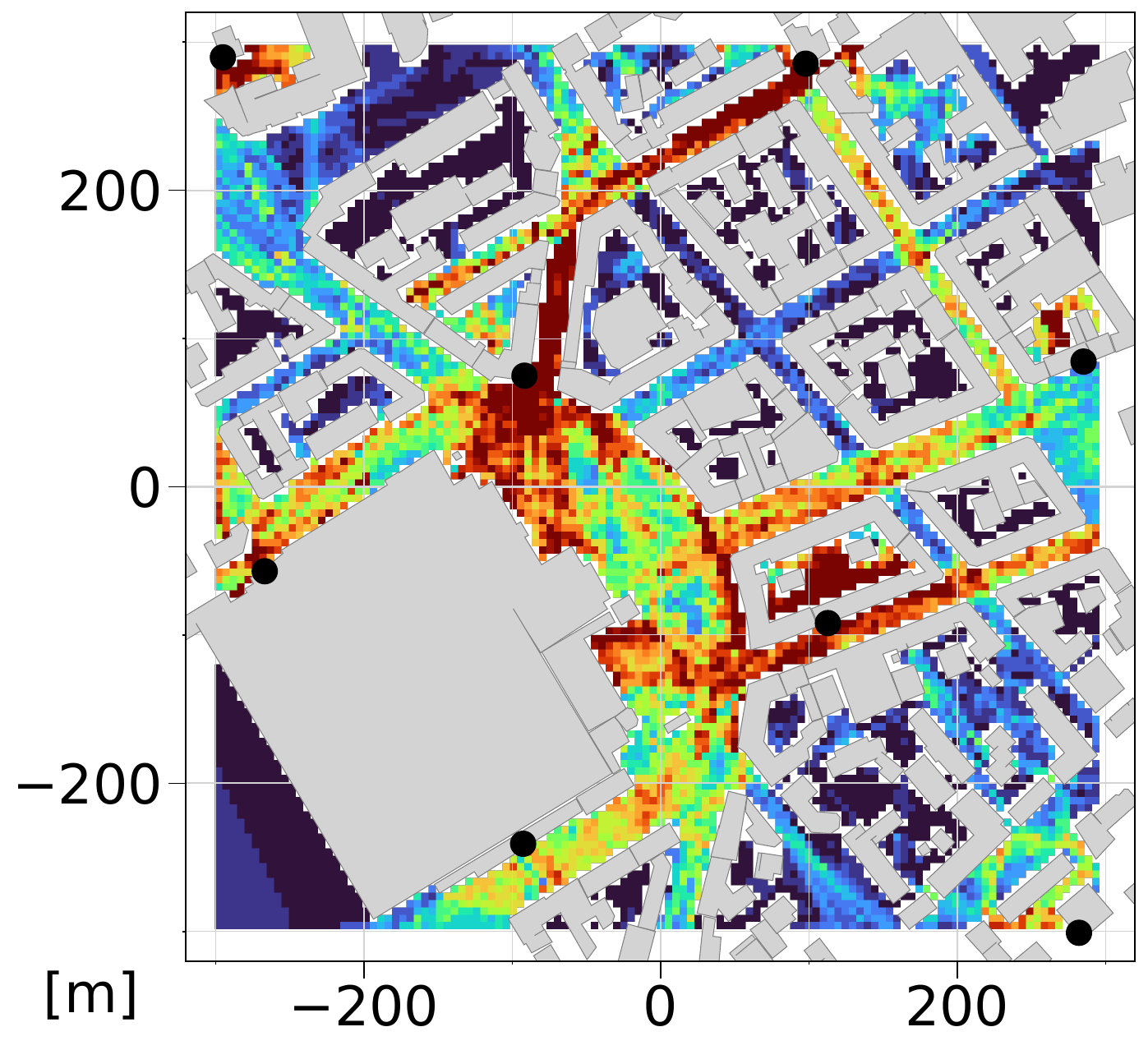}
            \caption{$f_c=8$ GHz}
            \label{fig:nb_paths_8_-5}
        \end{minipage}
\hfill
\begin{minipage}{0.51\textwidth}
            \includegraphics[width=\textwidth]{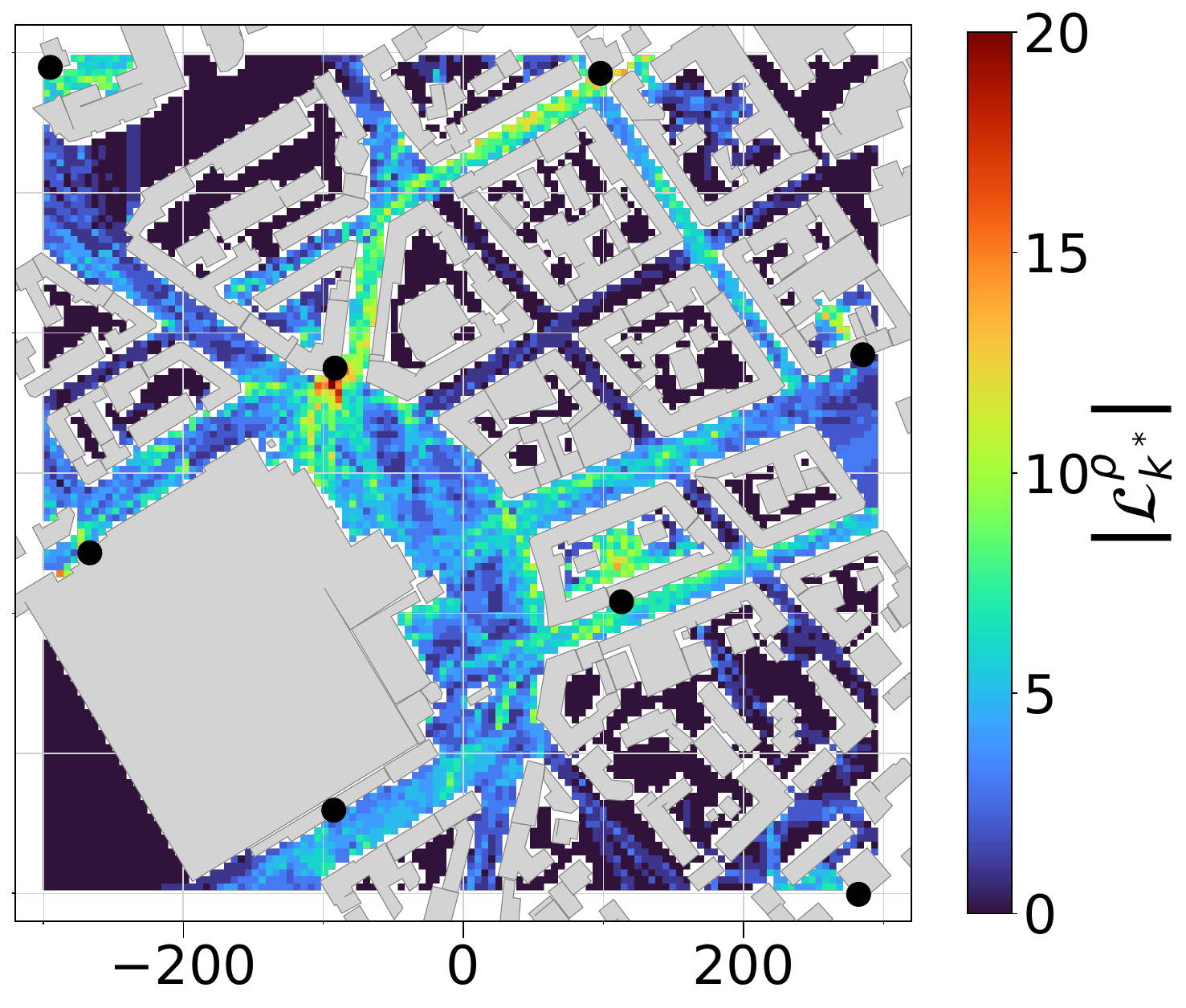}
            \caption{$f_c=18$ GHz}
            \label{fig:nb_paths_18_-5}
        \end{minipage}
    \end{subfigure}
    \caption{Heatmap of the number of dominant propagation paths for different FR3 bands (\emph{cf.} Footnote 2).}
    \label{fig:nb_paths}
\end{minipage}
\vspace{-3mm}
\end{figure}

Finally, we note that the decreasing proportion of valid beams with increasing frequency observed in Fig.~\ref{fig:ccdf_valid_beams_comparison} is consistent with the increase in channel sparsity with increasing frequency \cite{shakya2025angular}. However, we emphasize that the difference in the proportion of valid beams among different frequency bands is significantly less pronounced than the difference in their multipath richness, i.e. channel sparsity. This is nicely illustrated by comparing the spatial distribution of the number of valid beams satisfying the SINR outage threshold of $-5$~dB for gNB 3 in Fig.~\ref{fig:valid_beam_spatial} with the spatial distribution of the number of dominant propagation paths at a given UE location, as given by the ray-tracer\footnote{We consider a ray-tracing propagation path as dominant if its signal strength is above the outage SNR of -5~dB, thereby approximately illustrating the number of strong multipath components without beamforming gain effects.}, in Fig.~\ref{fig:nb_paths}. For example, the UE locations around the building corner where gNB 3 is located (i.e. directly below the gNB and up the street to the right of it) have a significantly higher number of dominant propagation paths at 8 GHz than 18 GHz, i.e. around 15-20 and 5-12 in Fig.~\ref{fig:nb_paths_8_-5} and \ref{fig:nb_paths_18_-5}, respectively. However, the proportion of valid beams at these UE locations is much more comparable at the two FR3 frequencies, as illustrated by the similar colours in this region of the heatmaps of Fig.~\ref{fig:valid_beam_8_-5} and \ref{fig:valid_beam_18_-5}. This less pronounced difference in directional link availability between FR3 bands in Fig.~\ref{fig:valid_beam_spatial} than suggested by the channel sparsity difference illustrated in Fig.~\ref{fig:nb_paths} is due to two main effects: (i) the higher directional beamforming gain of a larger antenna array at the serving gNB, which boosts weaker multipath components at higher frequencies; and (ii) the interference from other gNBs being weaker at higher frequencies.

\subsection{Beam Management Effort at FR3 in a Mobility Scenario}
\label{sec:mobility_case_study}

\begin{figure}
 \centering
        \begin{subfigure}{0.48\textwidth}
        \centering
        \includegraphics[width=\linewidth]{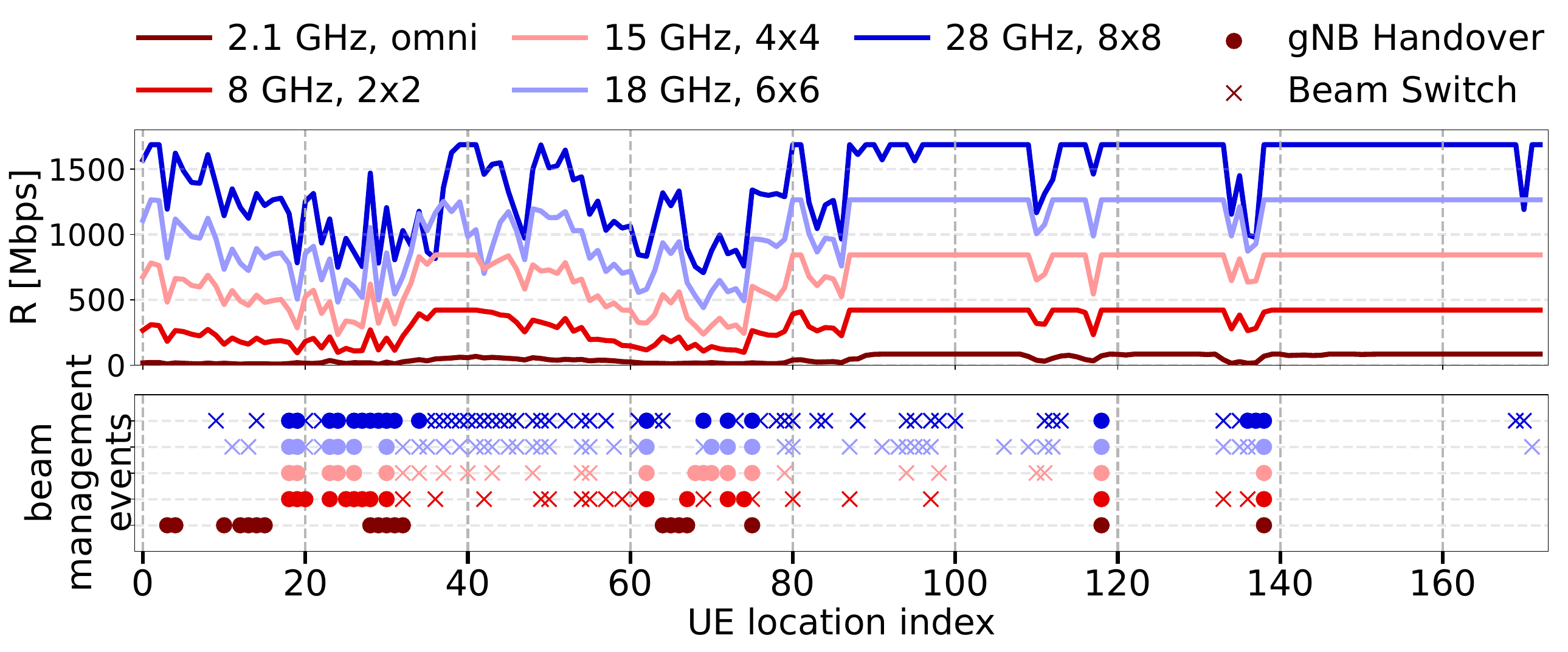}
        \caption{$\rho=\rho_{max}$}
        \label{fig:threshold_optimal}
    \end{subfigure}
\\
    \begin{subfigure}{0.48\textwidth}
        \centering
        \includegraphics[width=\linewidth]{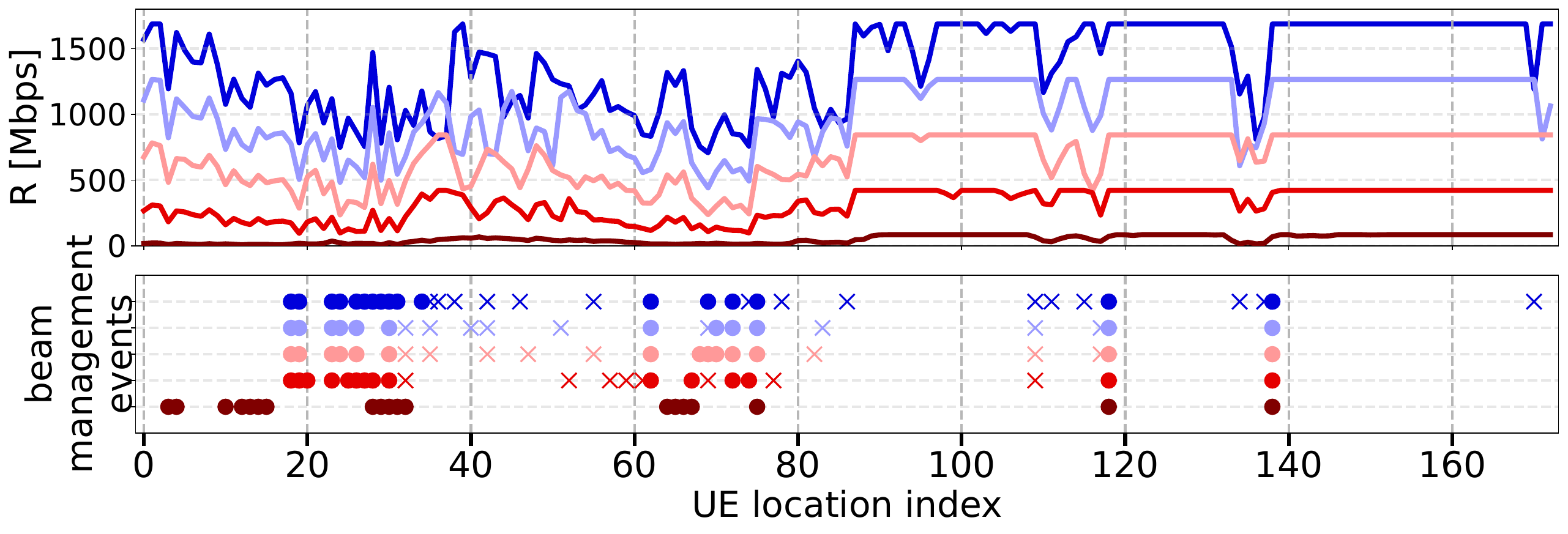}
        \caption{$\rho = 10$~dB}
        \label{fig:threshold10}
    \end{subfigure}
\\
    \begin{subfigure}{0.48\textwidth}
        \centering
        \includegraphics[width=\linewidth]{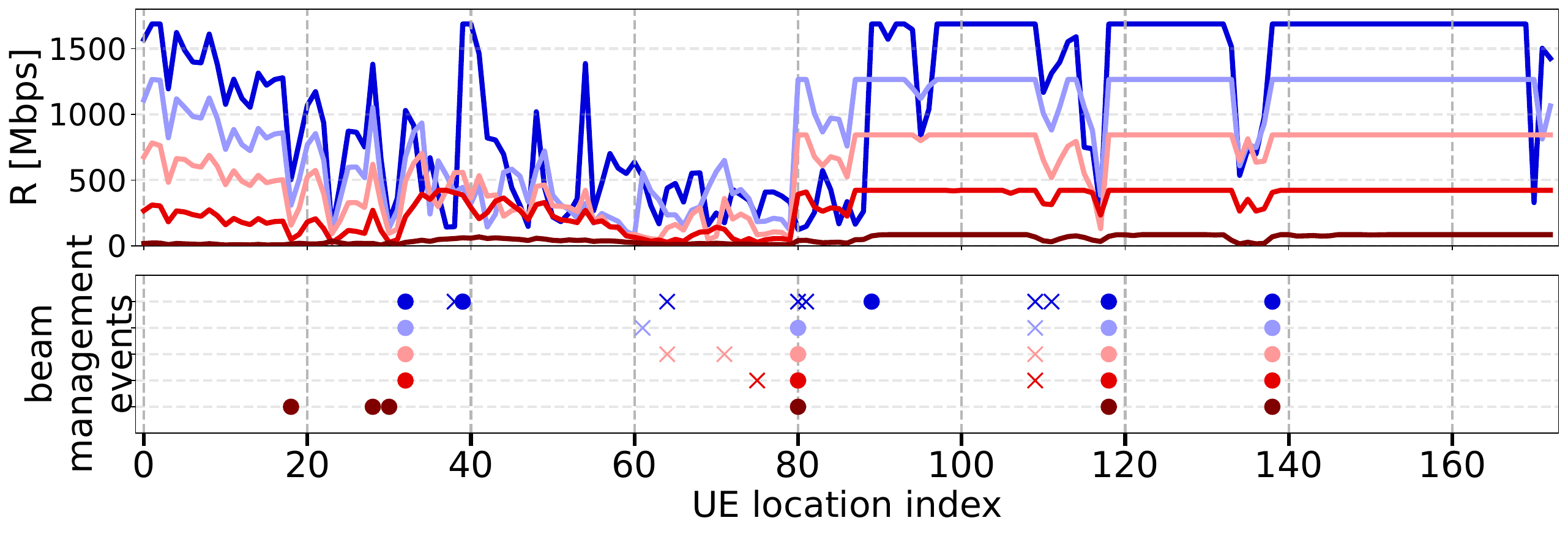}
        \caption{$\rho = -5$~dB}
        \label{fig:threshold-5}
    \end{subfigure}
    \caption{Data rate and beam management events (gNB handover or beam switch) for a mobile UE along the path in Fig.~\ref{fig:example_path}, for different frequency bands and SINR thresholds $\rho$.}
    \label{fig:data_rate_along_path}
    \vspace{-4mm}
\end{figure}

\begin{figure}
 \centering
    \begin{subfigure}{0.48\textwidth}
        \centering
        \includegraphics[width=\linewidth]{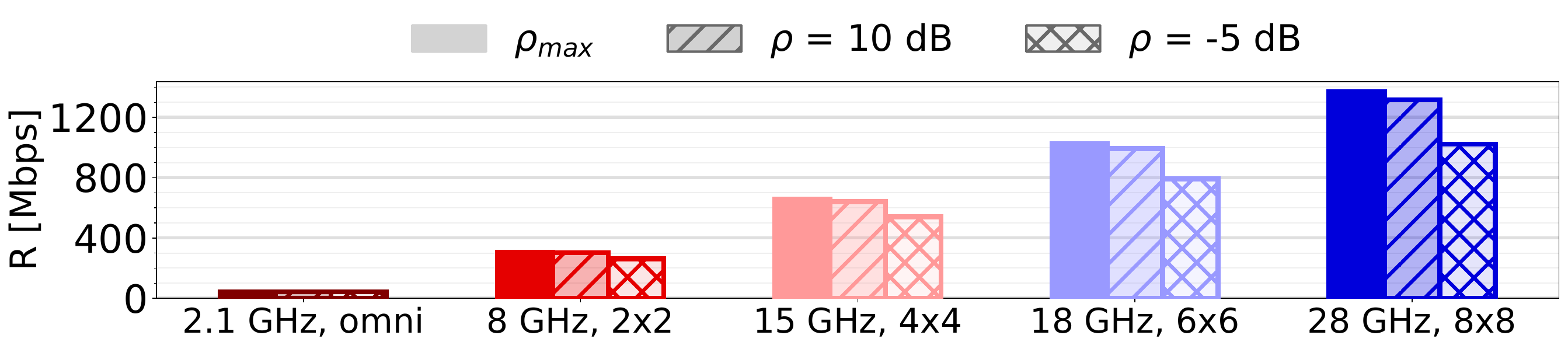}
        \caption{average data rate}
        \label{fig:barplot_dr}
    \end{subfigure}
\\
    \begin{subfigure}{0.48\textwidth}
        \centering
        \includegraphics[width=\linewidth]{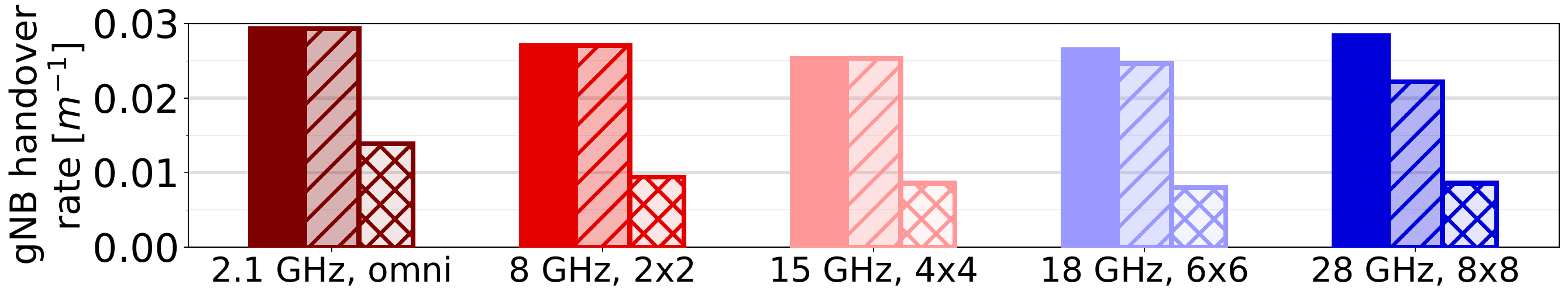}
        \caption{average gNB handover rate}
        \label{fig:barplot_handover}
    \end{subfigure}
\\
    \begin{subfigure}{0.48\textwidth}
        \centering
        \includegraphics[width=\linewidth]{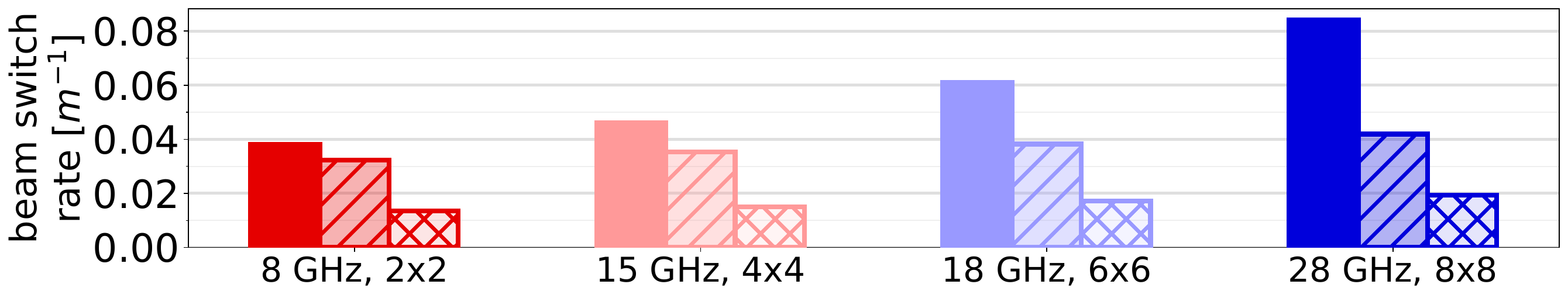}
        \caption{average beam switch rate}
        \label{fig:barplot_beam}
    \end{subfigure}
\\
\begin{subfigure}{0.48\textwidth}
        \centering
        \includegraphics[width=\linewidth]{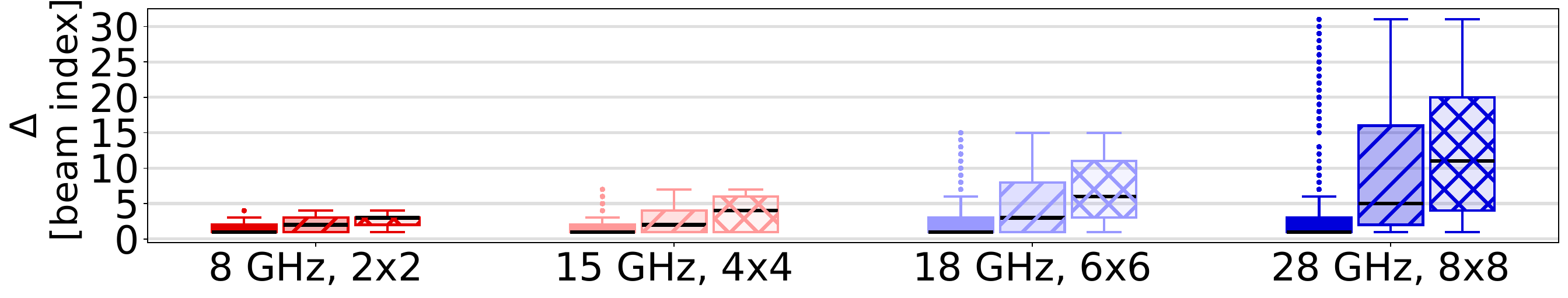}
        \caption{distribution of steering distance $\Delta(c)$ per beam switch}
        \label{fig:boxplot_steering_distance}
    \end{subfigure}
    \caption{Mobility statistics over the $M=2000$ UE paths in Fig.~\ref{fig:viswalk_paths}, for different frequency bands and SINR thresholds $\rho$.}
    \label{fig:statistics}
    \vspace{-3mm}
\end{figure}

Lastly, let us directly characterize the beam management effort at FR3 for a mobile UE. We assume a simple beam selection policy that approximates the minimum beam management effort to maintain a target SINR performance. The UE associates to its best gNB and optimal beam at the start of its mobility path and then for each subsequent UE position: (i) the UE stays on the same beam if the SINR threshold $\rho$ is maintained; (ii) otherwise, a \emph{beam switch} to the best beam on the serving gNB is performed, if this satisfies $\rho$; (iii) otherwise, we select the best beam over all gNBs thus potentially performing a beam switch (on the serving gNB) or a \emph{gNB handover}. Fig.~\ref{fig:data_rate_along_path} presents the resulting data rate along the example UE path in Fig.~\ref{fig:example_path} and records the beam management events (i.e. handover or beam switch), for different frequencies and target SINRs $\rho = \{-5, 10, \rho_{max}\}$~dB. Fig.~\ref{fig:statistics} presents the corresponding data rate and beam management event statistics over the $M=2000$ UE mobility paths in Fig.~\ref{fig:viswalk_paths}. Fig.~\ref{fig:statistics} additionally presents the distribution of the beam steering distance $\Delta_c$ in (\ref{eq:steering_distance}) per beam switch event.

Fig.~\ref{fig:threshold_optimal} represents the beam management effort to maximize the achievable data rate along the example UE path (since $\rho_{max}$ gives the $R_{max}$ per band, \emph{cf.} Table~\ref{tab:conductivity}). Fig.~\ref{fig:threshold_optimal} shows that the maximum achievable mobile data rate fluctuates significantly along the UE's path, regardless of frequency band, consistent with the varying achievable SINR coverage at different network locations (\emph{cf.} Fig.~\ref{fig:heatmap_sinr_opt_8ghz}).
Fig.~\ref{fig:threshold_optimal} also shows that the FR3 and FR2 bands exhibit a similar number of gNB handovers at similar UE positions, roughly corresponding to the best serving gNB distribution in Fig.~\ref{fig:heatmap_best_gNB_8ghz}.  FR1 exhibits some ping-pong handovers in areas with more challenging propagation such as the beginning of the UE's path, given the overall poorer SINR coverage at FR1 (\emph{cf.} Fig.~\ref{fig:ccdf_snr_90}). Nonetheless, the 
average handover rates in Fig.~\ref{fig:barplot_handover} are comparable across the frequency bands, i.e. around 0.029, 0.026, and 0.028 handovers/m at FR1, FR3, and FR2, respectively.

Importantly, Fig.~\ref{fig:threshold_optimal} shows that maximizing the data rate entails a significant number of beam switch events at both FR2 and FR3, corresponding to an average rate of \{0.04, 0.05, 0.06, 0.08\} beam-switches/m at \{8, 15, 18, 28\}~GHz, respectively, in Fig.~\ref{fig:barplot_beam}. Moreover, Fig.~\ref{fig:boxplot_steering_distance} shows that the corresponding median and upper-quartile beam steering distance per switch are 1 and 2-3, respectively, for all FR3 and FR2 bands.
This indicates that the higher beam switch rate for higher bands in Fig.~\ref{fig:threshold_optimal} is mainly due to more frequent but straightforward beam tracking to a near-adjacent beam.

Comparing Fig.~\ref{fig:threshold_optimal} with Figs.~\ref{fig:threshold10}~and~\ref{fig:threshold-5} shows that we can relax the required beam management effort at FR3 by lowering the threshold $\rho$. The resulting average FR3 mobile data rate is somewhat lower for lower $\rho$ thresholds, but Fig.~\ref{fig:barplot_dr} shows it is still over \{250, 550, 800\}~Mbps across the FR3 bands and thus remains superior to the average FR1 mobile rate of under 50~Mbps. However, this relaxed beam management requirement comes at the cost of a much more unstable FR3 mobile data rate. For example, in Fig.~\ref{fig:threshold-5} we allow the SINR (and thus rate) to deteriorate on a misaligned beam up to the outage threshold and only then perform a beam switch to the best available beam. Consequently, these remaining unavoidable beam switch events in Fig.~\ref{fig:threshold-5} typically have a larger steering distance, i.e. median of \{3, 4, 6, 11\} beams at \{8, 15, 18, 28\}~GHz, respectively, in Fig.~\ref{fig:boxplot_steering_distance}, suggesting they trigger more expensive (in terms of delay overhead) beam-sweep search procedures than when switching to adjacent beams. 
The results for $\rho$=-5~dB in Fig.~\ref{fig:barplot_beam} also indicate that the rate of incurring such a non-adjacent beam switch is only 12-30\% less at FR3 than FR2.

\section{Conclusions}
\label{sec:ccl}
We presented the first quantitative evaluation of beam management effort across the ``golden'' 6G FR3 frequency bands, considering realistic UE mobility and urban propagation via ray-tracing in an FR1-like cellular network deployment. Our results  confirm that the spectrum-rich FR3 bands can easily outperform FR1 mobile data rates. However, achieving both high and stable mobile throughput requires significant beam management effort across  all FR3 frequency bands, similar to FR2. Indeed, we show that the difference in directional link opportunities between frequency bands is less pronounced than the difference in their channel sparsity, due to higher beamforming gain and weaker interference at higher frequencies. Consequently, the rate of beam tracking for a mobile UE is more relaxed at lower frequencies (employing smaller antenna arrays with wider beams), but high-overhead beam switching to a non-adjacent beam is required at a relatively comparable rate for FR3 and FR2. Our findings motivate developing beam management strategies adaptive to the frequency band for spectrum-agile 6G networks.

\bibliography{references}

\end{document}